\documentclass{article}
\usepackage{amsmath,graphicx}
\usepackage{xcolor}

\usepackage{multirow}
\usepackage{makecell}
\usepackage{enumitem} % itemize left margin

\usepackage[preprint]{spconf}
\copyrightnotice{\copyright IEEE 2024}
\toappear{Submitted to {\it ICASSP 2024, April 14-19, 2024, Seoul, South Korea}}

\title{Multilingual Audio-Visual Speech Recognition with Hybrid CTC/RNN-T Fast Conformer}

\name{Maxime Burchi$^{1,2}$\sthanks{Work done during an internship at NVIDIA}, Krishna C. Puvvada$^{1}$, Jagadeesh Balam$^{1}$, Boris Ginsburg$^{1}$, Radu Timofte$^{2}$}
\address{$^{1}$NVIDIA, USA \\ $^{2}$Computer Vision Lab, CAIDAS \& IFI, University of W\"urzburg, Germany}

\begin{document}
%\ninept
%
\maketitle
\begin{abstract}

Humans are adept at leveraging visual cues from lip movements for recognizing speech in adverse listening conditions. Audio-Visual Speech Recognition (AVSR) models follow similar approach to achieve robust speech recognition in noisy conditions. In this work, we present a multilingual AVSR model incorporating several enhancements to improve performance and audio noise robustness. Notably, we adapt the recently proposed Fast Conformer model to process both audio and visual modalities using a novel hybrid CTC/RNN-T architecture. We increase the amount of audio-visual training data for six distinct languages, generating automatic transcriptions of unlabelled multilingual datasets (VoxCeleb2 and AVSpeech). Our proposed model achieves new state-of-the-art performance on the LRS3 dataset, reaching WER of 0.8\%. On the recently introduced MuAViC benchmark, our model yields an absolute average-WER reduction of 11.9\% in comparison to the original baseline. Finally, we demonstrate the ability of the proposed model to perform audio-only, visual-only, and audio-visual speech recognition at test time.

\end{abstract}
\section{Introduction}

Audio-Visual Speech Recognition (AVSR) has attracted a lot of research attention in recent years due to its ability to use image processing techniques to aid speech recognition systems. Preceding works have shown that including the visual modality of lip movements could improve the robustness of ASR systems with respect to audio noise while achieving better recognition performance~\cite{son2017lip, sterpu2018attention, afouras2018deep, petridis2018audio, xu2020discriminative, ma2021end, burchi2023audio}. Son~\textit{et al.}~\cite{son2017lip} proposed to use the sequence-to-sequence encoder-decoder with attention architecture to recognise phrases and sentences being spoken by talking faces, with or without the audio. Petridis~\textit{et al.}~\cite{petridis2018audio} applied the hybrid CTC/attention architecture to AVSR, achieving better performance and robustness using an early fusion strategy. Ma~\textit{et al.}~\cite{ma2021end} proposed to encode audio and visual modalities with Conformer~\cite{gulati2020conformer} networks to model both local and global temporal relationships using convolution and attention. More recently, AVEC~\cite{burchi2023audio} added intermediate CTC losses~\cite{nozaki2021relaxing} in encoder networks to improve lip reading performance and AVSR robustness.

Supervised AVSR approaches have shown promising results with stronger audio noise robustness but still rely on a limited amount of human-annotated speech data compared to classical ASR systems. This is especially the case for non-English languages, for which the amount of available data is not as important. Serdyuk~\textit{et al.}~\cite{serdyuk2022transformer} and Chang~\textit{et al.}~\cite{chang2023conformers} showed that performance could greatly be improved using large scale datasets of up to 100k hours composed of YouTube videos. However, these datasets remain private and unusable for comparison. LRS3~\cite{afouras2018lrs3}, which is the largest publicly available AVSR dataset, contains less than 450 training hours of audio-visual speech while LRW~\cite{chung2017lip} and LRS2~\cite{afouras2018deep} datasets are only available to academic institutions. The MuAViC corpus~\cite{anwar2023muavic}, which was recently released as benchmark for multilingual AVSR, contains only 10 hours of German speech.

To remedy this problem, other works started to focus on using self-supervised pre-training techniques~\cite{shi2022learning, haliassos2022jointly, hsu2022u, lian2023av, zhu2023vatlm} to improve performance by learning hidden representations with large scale unlabeled datasets. AV-HuBERT~\cite{shi2022learning} was the first self-supervised system to jointly learn speech representations from audio and video. It iteratively learns to minimize a masked prediction loss by first clustering acoustic features (MFCC) and then audio-visual hidden features using k-means. RAVen~\cite{haliassos2022jointly} proposed to pre-train unimodal encoders by predicting cross-modal representations of slow-moving teacher networks. VATLM~\cite{zhu2023vatlm} proposed a visual-audio-text model optimized to predict the hidden units of different modalities with a unified masked prediction task.\iffalse u-HuBERT~\cite{hsu2022u} generalizes AV-HuBERT to utilize both multimodal and unimodal data using modality dropout during pre-training.\fi~AV-data2vec~\cite{lian2023av} learns to predict the unmasked audio-visual output representations of a slow-moving teacher network. All these methods successfully improve generalization and performance by pre-training on the VoxCeleb2~\cite{chung2018voxceleb2} dataset with various pretext tasks. However, they can be complex to pre-train and still require to be fine-tuned.

More recently, Ma~\textit{et al.}~\cite{ma2023auto} showed that state-of-the-art performance could simply be achieved by training on generated transcriptions. They first generate English transcriptions of VoxCeleb2 and AVSpeech datasets using Whisper~\cite{radford2023robust} and use them to increase the amount of training data available. Inspired by this work, we study the impact of generated transcriptions on six languages using large scale multilingual datasets. This paper brings four main contributions: 
\vspace{-0.1cm}
\begin{itemize}[leftmargin=0.5cm]
    \setlength\itemsep{-0.2em}
    \item A novel hybrid architecture for AVSR using both Connectionist Temporal Classification (CTC)~\cite{graves2006connectionist} and Recurrent Neural Network Transducer (RNN-T)~\cite{graves2012sequence} losses. 
    \item The use of generated transcriptions from large scale multilingual datasets (VoxCeleb2 and AVSpeech) to improve non-English AVSR performance and robustness. 
    \item New state-of-the-art results on the LRS3~\cite{afouras2018lrs3} dataset and the recently introduced MuAViC~\cite{anwar2023muavic} benchmark.
    \item The ability to use a single model for audio-only, visual-only and audio-visual speech recognition at test time using modality masking before audio-visual fusion. 
\end{itemize}

\begin{figure}[tb]
        \centering
        \includegraphics[width=0.95\linewidth]{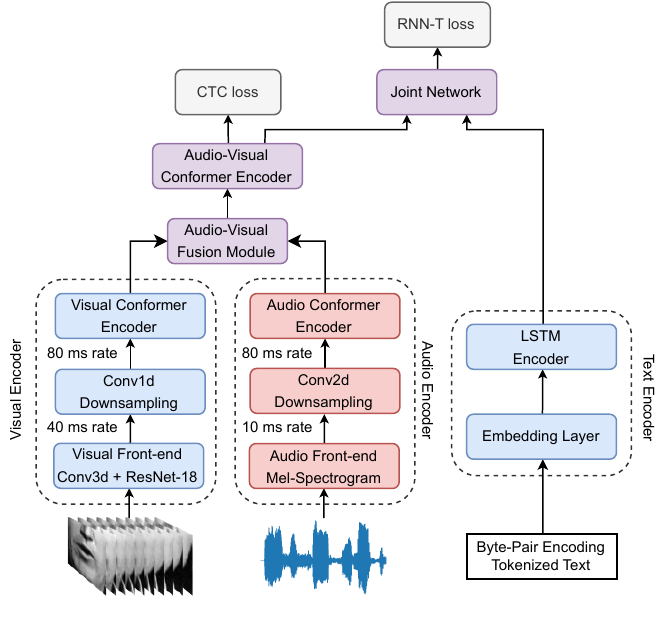}
        \vspace{-0.5cm}
        \caption{\textbf{Audio-Visual Fast Conformer architecture.} The model is trained end-to-end using CTC and RNN-T losses and takes both audio waveforms and lip movements as inputs.}
    \label{fig:model}
    \vspace{-0.5cm}
\end{figure}

\vspace{-0.35cm}
\section{Model Architecture}

Our model design is based on the recently proposed Fast Conformer ASR architecture~\cite{rekesh2023fast}. The Fast Conformer brings compute-memory savings compared to Conformer by further downsampling the input audio mel-spectrogram by a factor of 2. It increases the sampling period from 10ms to 80ms using 3 depth-wise convolutional sub-sampling layers. We adapt this architecture to AVSR by processing both audio and visual inputs, adding a ResNet-based front-end network for visual pre-processing and adopting an early fusion strategy~\cite{petridis2018audio}. We train our models using both CTC and RNN-Transducer losses, disposing of the need to train both models separately while improving RNN-T performance. The model has 18 Conformer encoder blocks with 512 hidden dimensions and a single layer LSTM~\cite{hochreiter1997long} decoder with 640 hidden dimensions. For audio-visual models, each modality is first processed separately by 10 Conformer blocks using unimodal encoders while 8 blocks are dedicated to the audio-visual encoder. This results in a total of 119M, 130M and 197M parameters for the audio-only, visual-only and audio-visual models, respectively.  The complete audio-visual model architecture is shown in Figure~\ref{fig:model}. 

\vspace{-0.25cm}
\subsection{Visual Front-End}

Similar to preceding works, we process input videos using a modified ResNet-18~\cite{he2016deep,ma2021end} where the first layer is a spatio-temporal convolutional layer with kernel size $5\times7\times7$. The resulting feature maps are average-pooled along spatial dimensions and downsampled using a strided temporal convolution layer. The resulting signal has a 80ms sampling period.

\vspace{-0.25cm}
\subsection{Modality Fusion}

We use an early fusion strategy~\cite{petridis2018audio} to learn audio-visual features and reduce model complexity. The acoustic and visual features are concatenated and fed into a joint feed-forward network. We also apply a modality dropout~\cite{shi2022learning} scheme during training to counter a common failure case where audio-visual models tend to ignore the visual modality~\cite{afouras2018deep}. Modality dropout is performed by masking one of the two modalities 30\% of the time before fusion. This requires the network to learn from both modalities while allowing unimodal inference at test time by masking one of the two modality signals.

\vspace{-0.25cm}
\subsection{Hybrid CTC/RNN-T loss}

The network is trained end-to-end using the RNN-Transducer objective with an additional CTC loss, as follows:
\vspace{-0.1cm}
\begin{align}
Loss = (1- \alpha) L_{RNNT} + \alpha L_{CTC}
\end{align}
Where $\alpha$ is a loss scaling constant that we fix to 0.3 in every experiment. Table~\ref{table:hybrid} compares the final performance of CTC, RNN-Transducer and hybrid networks on the LRS3 test set.

\section{Experiments}

\subsection{Datasets}

We conduct experiments on two publicly available audio-visual datasets, both extracted from TED and TEDx talks: LRS3~\cite{afouras2018lrs3}, which is an English only AVSR dataset, and MuAViC~\cite{anwar2023muavic}, which is multilingual. LRS3 is the largest publicly available English AVSR dataset, including 438.9 hours of audio-visual speech. It is composed of a pre-training set (408 hours), a training-validation set (30 hours) and a test set (0.9 hours) for evaluation. The MuAViC dataset was recently introduced as a multilingual audio-visual corpus. The corpus includes 1200 hours of transcribed audio-visual speech from over 8k speakers in 9 languages: English (en), Arabic (ar), German (de), Greek (el), Spanish (es), French (fr), Italian (it), Portuguese (pt) and Russian (ru). Where the English subset is the addition of LRS3 pre-training and training-validation sets, which we refer to as LRS3 throughout the paper. In this work, we concentrate on Romance and Germanic language subsets for English, Spanish, French, Portuguese Italian and German. We extract additional audio-visual speech data from the large scale VoxCeleb2 (VC2)~\cite{chung2018voxceleb2} and AVSpeech (AVS)~\cite{ephrat2018looking} datasets to increase the amount of training data per language. VC2 contains approximately 2400 hours of videos extracted from YouTube with a million utterances by 6000+ celebrities, while AVSpeech has 4700 hours from 150k speakers. Both datasets include a wide variety of people, languages, and face poses. 
We detect language ids and generate transcriptions for each audio segment using Whisper Large~\cite{radford2023robust}, a 1.55B parameters multilingual model trained on 680k hours of speech data. For English segments, we use a 1.1B parameter Fast Conformer XXL~\cite{rekesh2023fast} model pre-trained on LibriLight and fine-tuned on 24.5k hours of English speech. Table~\ref{table:datasets} shows the total amount of training hours per dataset and language. 

\begin{table}[ht]
\vspace{-0.5cm}
\centering
\setlength{\tabcolsep}{6pt}
\scriptsize
\caption{Number of audio-visual speech hours.}
\hfill \break
\begin{tabular}{l|ccccccc}
\hline
\multirow{2}{*}{\textbf{Dataset}} & \multicolumn{7}{c}{\textbf{Number of training hours per language}}\\ %\cline{5-7}
& \textbf{en} & \textbf{es} & \textbf{fr} & \textbf{pt} & \textbf{it} & \textbf{de} & \textbf{all}\\
\hline\hline
LRS3 / MuAViC & 435 & 176 & 173 & 152 & 99 & 10 & 1046 \\
VoxCeleb2 (generated) & 1252 & 37 & 109 & 8 & 36 & 153 & 1595 \\
AVSpeech (generated) & 1429 & 257 & 115 & 312 & 66 & 136 & 2316 \\
\hline
Total generated & 2681 & 294 & 224 & 319 & 102 & 289 & 3910 \\
Total & 3116 & 470 & 398 & 472 & 201 & 299 & 4957 \\
\hline
\end{tabular}
\label{table:datasets}
\vspace{-0.5cm}
\end{table}

\subsection{Implementation details}
For visual pre-processing, we follow previous works \cite{ma2021end,burchi2023audio} to prepare all the datasets. The RetinaFace~\cite{deng2020retinaface} face detector is used to detect speaker faces. We then remove differences related to rotation and scale by cropping speaker lip regions using bounding boxes of $96 \times 96$ pixels to facilitate recognition.  We localize 68 face landmarks using a Face Alignment Network~\cite{bulat2017far} to align the speaker face and crop lip regions. The cropped images are then converted to grayscale.

For data augmentation, we apply Spec-Augment~\cite{park2020specaugment} to audio mel-spectrograms with mask size parameter $F = 27$ and ten time masks with adaptive size $pS = 0.05$. We also add babble noise from the NoiseX corpus~\cite{varga1993noisex} during training, as done in previous works~\cite{ma2021end, burchi2023audio}. On the visual side, we apply random cropping with crop size $88\times88$, horizontal flipping and time masking~\cite{ma2022visual}. We apply center crop at test time.

We first train audio-visual, audio-only and visual-only models on the LRS3 dataset augmented with Voxceleb2 and AVSpeech English transcriptions. Ma~\textit{et al.}~\cite{ma2023auto} pre-train the visual-only models on the LRW dataset using a word classification task to facilitate early convergence and improve final performance. However, since we do not have access to this dataset, we first train the audio-visual model and use the ResNet-18 weights as initialization for the visual-only model. We also add intermediate CTC losses~\cite{lee2021intermediate} every 3 Conformer blocks to the visual-only model encoder to speed up convergence. Training is done using the AdamW~\cite{loshchilov2017decoupled} optimizer with $\beta_{1} = 0.9$, $\beta_{2} = 0.98$ and a 1e-3 weight decay. The models are trained for 100k gradient steps with a global batch size of 2048 samples. The learning rate is ramped up linearly to 0.001 during the first 5k steps, held constant until 15k steps and decayed following an inverse square root schedule. An Exponential Moving Average (EMA) model with 0.9999 momentum is updated along training for evaluation.

For MuAViC, we train monolingual and multilingual models with encoder weights initialized from English pre-trained networks. We use a SentencePiece~\cite{kudo2018sentencepiece} byte-pair-encoding tokenizer with 256 tokens per language. Multilingual tokenization is done using an aggregated tokenizer~\cite{dhawan2023towards}, where each language is given a range of tokens and the corresponding range is used during decoding. Multilingual models are trained for 100k steps while monolingual models are trained until convergence, for no more than 1000 epochs.

\vspace{-0.15cm}
\subsection{Results on LRS3}

Table~\ref{table:results_lrs3} compares the Word Error Rates (WER) of our English models with recently published methods on the LRS3 test set. Our audio-only and audio-visual models achieve new state-of-the-art results with WER of 0.7\% and 0.8\%, respectively. Our lip reading network competes with self-supervised approaches with 25.5\% WER. However, it still lags behind Auto-AVSR~\cite{ma2023auto} whose visual-only model has 250.4M parameters, trains on additional LRS2 human-labeled transcriptions and uses an external language model during decoding. 

\begin{table}[ht]
\vspace{-0.4cm}
\centering
\setlength{\tabcolsep}{3.5pt}
\scriptsize
\caption{Comparison of WER (\%) on the LRS3 test set with recently published methods for Audio-Visual (AV), Audio-Only (A) and Visual-Only (V) speech recognition models. $^{\ast}$~Shows number of encoder params instead of total params. $^{\ddagger}$~Use of additional audio, audio-text and text data~\cite{zhu2023vatlm, hsu2022u}.}
\hfill \break
\begin{tabular}{l|clcccc}
\hline
\multirow{2}{*}{\textbf{Method}} & \multicolumn{1}{c}{\multirow{2}{*}{\begin{tabular}[c]{@{}c@{}}\textbf{AV model} \\ \textbf{Params}\end{tabular}}} & \multicolumn{1}{c}{\multirow{2}{*}{\begin{tabular}[c]{@{}c@{}}\textbf{Training} \\ \textbf{Datasets}\end{tabular}}} & \multicolumn{1}{c}{\multirow{2}{*}{\begin{tabular}[c]{@{}c@{}}\textbf{Total} \\ \textbf{Hours}\end{tabular}}} & \multicolumn{3}{c}{\textbf{LRS3 test WER}}\\ %\cline{5-7}
& & & & \textbf{AV} &  \textbf{A} &  \textbf{V} \\
\hline\hline
CM-seq2seq~\cite{ma2021end} & 92M & LRW,LRS3 & 595 & 2.3 & 2.3 & 43.3 \\
AVEC~\cite{burchi2023audio} & 61M & LRW,LRS2\&3 & 818 & 1.8 & 2.0 & 37.5  \\

AV-HuBERT~\cite{shi2022learning, lian2023av} & 103M$^{\ast}$ & LRS3,VC2 & 1759 & 1.8 & 2.0 & 34.8 \\
RAVen~\cite{haliassos2022jointly} & 41M$^{\ast}$ & LRS3,VC2 & 1759 & - & 1.9 & 33.1 \\
VATLM~\cite{zhu2023vatlm}$^{\ddagger}$ & 103M$^{\ast}$ & LRS3,VC2 & 1759 & 1.7 & - & 34.2 \\
AV-data2vec~\cite{lian2023av} & 103M$^{\ast}$ & LRS3,VC2 & 1759 & 1.4 & 1.7 & 32.9 \\

AV-HuBERT~\cite{shi2022learning, shi2022robust} & 325M$^{\ast}$ & LRS3,VC2 & 1759 & 1.4 & 1.3 & 28.6 \\
RAVen~\cite{haliassos2022jointly} & 328M$^{\ast}$ & LRS3,VC2 & 1759 & - & 1.4 & 28.2 \\
VATLM~\cite{zhu2023vatlm}$^{\ddagger}$ & 325M$^{\ast}$ & LRS3,VC2 & 1759 & 1.2 & - & 28.4 \\
u-HuBERT~\cite{hsu2022u}$^{\ddagger}$ & 325M$^{\ast}$ & LRS3,VC2 & 1759 & 1.2 & 1.4 & 27.2 \\
AV-data2vec~\cite{lian2023av} & 325M$^{\ast}$ & LRS3,VC2 & 1759 & 1.3 & 1.4 & 28.5 \\

Auto-AVSR~\cite{ma2023auto} & 425M & +LRW,LRS2,AVS & 3448 & 0.9 & 1.0 & 19.1  \\
VIT 3D~\cite{serdyuk2022transformer} & 310M & YouTube-90k & 90k & 1.6 & - & 17.0 \\
VGG Conformer~\cite{chang2023robustness} & 180M & YouTube-100k & 100k & 0.9 & 1.0 & - \\
LP Conformer~\cite{chang2023conformers} & 570M & YouTube-100k & 100k & 0.9 & - & \textbf{12.8} \\
\hline
 \multirow{3}{*}{Fast Conformer} &  & LRS3 & 435 & 1.7 & 1.6 & 43.8\\
 & 197M & LRS3,VC2 & 1687 & 0.9 & \textbf{0.7} & 31.0  \\
 &  & LRS3,VC2,AVS & 3116 & \textbf{0.8} & \textbf{0.7} & 25.5 \\

\hline
\end{tabular}
\label{table:results_lrs3}
\vspace{-0.1cm}
\end{table}

We evaluate our English models robustness, adding babble noise and white noise with multiple Signal to Noise Ratio (SNR) during test time. Table~\ref{table:lr3_robustness} compares the robustness of our model with Auto-AVSR on the LRS3 test set. We observe that our models can recover similar audio robustness by training on generated transcriptions while using half the number of training parameters. It also shows that audio-only models can easily overfit to babble noise, demonstrating the importance of evaluating on noises not encountered during training.

\iffalse
\begin{table}[ht]
\vspace{-0.25cm}
\centering
\setlength{\tabcolsep}{4pt}
\scriptsize
\caption{Audio noise robustness on LRS3 test set.}
\hfill \break
\begin{tabular}{c|cccccccc}
\hline
\multirow{2}{*}{\textbf{Model}} & \multirow{2}{*}{\textbf{Mode}} & \multicolumn{1}{c}{\multirow{2}{*}{\begin{tabular}[c]{@{}c@{}}\textbf{Num} \\ \textbf{Params}\end{tabular}}} & \multirow{2}{*}{\textbf{Noise}} & \multicolumn{5}{c}{\textbf{test time SNR level (dB)}}\\ 
& & & & \textbf{12.5} & \textbf{7.5} & \textbf{2.5} & \textbf{-2.5} & \textbf{-7.5} \\
\hline\hline

 \multirow{4}{*}{Auto-AVSR~\cite{ma2023auto}} & A & 243M & \multirow{2}{*}{babble} & 1.1 & 1.2 & 1.6 & 2.7 & 8.3\\
 & AV & 425M & & 1.0 & 1.0 & 1.5 & 2.2 & 5.6 \\
  & A & 243M & \multirow{2}{*}{white} & 2.1 & 4.0 & 10.4 & 30.2 & 88.9 \\
 & AV & 425M & & 1.4 & 2.3 & 4.3 & 9.5 & 24.2 \\

\hline

 \multirow{4}{*}{Fast Conformer} & A & 119M & \multirow{2}{*}{babble} & 0.9 & 1.2 & 1.8 & 3.5 & 10.9\\
 & AV & 197M & & 1.3 & 1.5 & 2.0 & 3.2 & 6.7 \\
 & A & 119M &  \multirow{2}{*}{white} & 1.5 & 2.8 & 6.9 & 17.6 & 54.8 \\
 & AV & 197M & & 1.4 & 2.1 & 3.2 & 5.7 & 21.4 \\
 
\hline
\end{tabular}
\label{table:lr3_robustness}
\vspace{-0.25cm}
\end{table}
\fi

\begin{table}[ht]
\vspace{-0.25cm}
\centering
\setlength{\tabcolsep}{4pt}
\scriptsize
\caption{LRS3 test set WER (\%) under noisy conditions. \\ $^{\dagger}$ Noise type used during both training and test time.}
\hfill \break
\begin{tabular}{l|cccccccc}
\hline
\multirow{2}{*}{\textbf{Method}} & \multirow{2}{*}{\textbf{Mode}} & \multicolumn{1}{c}{\multirow{2}{*}{\begin{tabular}[c]{@{}c@{}}\textbf{Num} \\ \textbf{Params}\end{tabular}}} & \multicolumn{1}{c}{\multirow{2}{*}{\begin{tabular}[c]{@{}c@{}}\textbf{Audio} \\ \textbf{Noise}\end{tabular}}} & \multicolumn{5}{c}{\textbf{test time SNR level (dB)}}\\ 
& & & & \textbf{12.5} & \textbf{7.5} & \textbf{2.5} & \textbf{-2.5} & \textbf{-7.5} \\
\hline\hline

 \multirow{2}{*}{Auto-AVSR~\cite{ma2023auto}} & A & 243M & \multirow{2}{*}{babble$^{\dagger}$} & 1.1 & 1.2 & 1.6 & 2.7 & 8.3\\
 & AV & 425M & & 1.0 & 1.0 & 1.5 & 2.2 & 5.6 \\
  \hline
 \multirow{2}{*}{Fast Conformer} & A & 119M & \multirow{2}{*}{babble$^{\dagger}$} & 0.9 & 1.2 & 1.8 & 3.5 & 10.9\\
 & AV & 197M & & 1.2 & 1.5 & 2.0 & 3.2 & 6.7 \\
 \hline
 \multirow{2}{*}{Auto-AVSR~\cite{ma2023auto}} & A & 243M & \multirow{2}{*}{white} & 2.1 & 4.0 & 10.4 & 30.2 & 88.9 \\
 & AV & 425M & & 1.4 & 2.3 & 4.3 & 9.5 & 24.2 \\
\hline
 \multirow{2}{*}{Fast Conformer} & A & 119M &  \multirow{2}{*}{white} & 1.5 & 2.8 & 6.9 & 17.6 & 54.8 \\
 & AV & 197M & & 1.3 & 2.1 & 3.2 & 5.7 & 21.4 \\
 
\hline
\end{tabular}
\label{table:lr3_robustness}
\vspace{-0.4cm}
\end{table}

\subsection{Results on MuAViC}

Table~\ref{table:muavic} compares our monolingual and multilingual models WER with MuAViC baseline models fine-tuned from AV-HuBERT~\cite{shi2022robust}. We also evaluate Whisper Large~\cite{radford2023robust}, which we use to generate transcriptions. We observe that our models trained solely on human-labeled transcriptions achieve better performance than the original work. Including additional generated transcriptions from VoxCeleb2 and AVSpeech datasets further improves performance for every language considered. This is especially the case for the German subset, which contains only 10 hours of human-labeled speech, improving audio-visual WER by almost 20\%. Moreover, we find that multilingual models can successfully be trained to recover similar performance by training on all languages, achieving an absolute average WER reduction of 11.94\% in comparison to the original baseline. Multilingual training also further improves robustness of audio-visual models, allowing the visual branch to learn from a wide variety of lip movements. Our model improves noisy WER by 28\% compared to Whisper, which was trained on 680k hours of multilingual speech data.

\begin{table}[ht]
\vspace{-0.4cm}
\centering
\setlength{\tabcolsep}{4.25pt}
\scriptsize
\caption{Comparison of WER (\%) on the MuAViC dataset.}
\hfill \break
\begin{tabular}{l|cccccccc}
\hline
\multirow{2}{*}{\textbf{Method}} & \multicolumn{1}{c}{\multirow{2}{*}{\begin{tabular}[c]{@{}c@{}}\textbf{Total} \\ \textbf{Hours}\end{tabular}}} & \multirow{2}{*}{\textbf{Mode}} & \multicolumn{5}{c}{\textbf{Language test WER}} & \multirow{2}{*}{\textbf{avg}} \\ %\cline{5-7}
& & & \textbf{es} & \textbf{fr} & \textbf{pt} & \textbf{it} & \textbf{de} \\
\hline\hline

Whisper~\cite{radford2023robust} (ours)& 680k & A & \textbf{7.6} & \textbf{9.4} & 11.3 & \textbf{8.7} & \textbf{20.4} & \textbf{11.5} \\
\hline

Monolingual & \multirow{2}{*}{2377} & A & 16.5 & 24.4 & 23.0 & 19.3 & 61.1 & 28.9\\
AV-HuBERT~\cite{anwar2023muavic} & & AV & 15.9 & 23.7 & 19.4 & 18.5 & 52.4 & 26.0\\
\hline

Multilingual & \multirow{2}{*}{2467} & A & 30.6 & 27.0 & 19.8 & 19.3 & 46.5 & 28.6\\
AV-HuBERT~\cite{anwar2023muavic} & & AV & 16.2 & 19.0 & 19.9 & 19.8 & 47.2 & 24.4\\
\hline

Monolingual & \multirow{2}{*}{3735} & A & 8.5 & 11.1 & 12.2 & 11.4 & 41.3 & 16.9\\
Fast Conformer & & AV & 9.4 & 11.4 & 12.7 & 12.4 & 43.9 & 18.0\\
%\hline
\cline{2-9}

~~~+ generated & \multirow{2}{*}{4957} & A & 7.9 & 9.6 & 10.2 & 10.2 & 21.7 & 11.9\\
~~~transcriptions & & AV & 8.2 & 9.8 & 10.2 & 10.6 & 22.4 & 12.2 \\
\hline

Multilingual & \multirow{2}{*}{3735} & A & 8.8 & 11.4 & 11.5 & 11.0 & 43.4 & 17.2\\
Fast Conformer & & AV & 9.0 & 11.4 & 11.8 & 11.6 & 46.8 & 18.1\\
%\hline
\cline{2-9}

~~~+ generated & \multirow{2}{*}{4957} & A & 8.2 & 10.6 & \textbf{9.9} & 10.2 & 23.4 & 12.5\\
~~~transcriptions & & AV & 8.2 & 10.3 & \textbf{9.9} & 10.4 & 23.6 & 12.5\\
 \hline\hline
 
\multicolumn{9}{c}{($\downarrow$) \textit{Adding White Noise with -5 SNR at test time} ($\downarrow$)}\\ 
 \hline

 Whisper~\cite{radford2023robust} (ours)& 680k & A & 56.3 & 61.3 & 72.4 & 59.4 & 69.2 & 63.7\\
\hline

Monolingual & \multirow{2}{*}{3735} & A & 58.7 & 57.8 & 63.0 & 61.5 & 91.4 & 66.5\\
Fast Conformer & & AV & 35.2 & 49.5 & 46.9 & 45.8 & 75.5 & 50.6 \\
\cline{2-9}

~~~+ generated & \multirow{2}{*}{4957} & A & 40.5 & 60.6 & 54.4 & 49.8 & 62.5 & 53.6\\
~~~transcriptions & & AV & 29.8 & 41.7 & 40.1 & 37.0 & \textbf{48.1} & 39.3\\
\hline

Multilingual & \multirow{2}{*}{3735} & A & 52.3 & 63.0 & 65.4 & 59.4 & 94.5 & 66.9\\
Fast Conformer & & AV & 35.1 & 41.9 & 42.4 & 37.5 & 87.0 & 48.8\\
%\hline
\cline{2-9}

~~~+ generated & \multirow{2}{*}{4957} & A & 44.9 & 52.9 & 55.9 & 50.0 & 69.0 & 54.5\\
~~~transcriptions & & AV & \textbf{28.2} & \textbf{35.1} & \textbf{34.7} & \textbf{30.3} & 50.3 & \textbf{35.7}\\
 \hline
 
\end{tabular}
\label{table:muavic}
\vspace{-0.65cm}
\end{table}

\subsection{Learning Objective}

We study the effect of the optimization architecture on final performance. Table~\ref{table:hybrid} compares the WER of CTC, RNN-T and hybrid models on the LRS3 test set. We find that adding a CTC loss slightly helps to improve RNN-T performance. Moreover, the hybrid CTC/RNN-T architecture can be used for CTC and RNN-T decoding without the need to train both networks separately. Following~\cite{burchi2023audio}, we also experiment adding intermediate CTC losses~\cite{lee2021intermediate} every 3 Conformer blocks and find it to further improve lip reading performance.

\begin{table}[ht]
\vspace{-0.4cm}
\centering
\scriptsize
\caption{Effect of loss type on Fast Conformer performance.}
\hfill \break
\begin{tabular}{l|lccc}
\hline
\multirow{2}{*}{\textbf{Loss Type}} & \multicolumn{1}{c}{\multirow{2}{*}{\begin{tabular}[c]{@{}c@{}}\textbf{Training} \\ \textbf{Datasets}\end{tabular}}} & \multicolumn{3}{c}{\textbf{CTC / RNN-T LRS3 test WER (\%)}}\\ %\cline{5-7}
& & \textbf{AV} & \textbf{A} & \textbf{V}\\
\hline\hline
CTC & LRS3 & 2.3 & 2.1 & 49.7 \\
RNN-T & LRS3 & 1.9 & 1.8 &  47.7 \\
CTC / RNN-T & LRS3 & 2.3 / 1.7 & 2.0 / 1.6 & 49.8 / 45.9 \\
InterCTC / RNN-T & LRS3 & 2.3 / 1.7 & 1.9 / 1.6 & 48.7 / 43.8 \\
CTC / RNN-T & LRS3,VC2,AVS & 1.4 / 0.8 & 1.1 / 0.7 & 31.1 / 27.0\\
%InterCTC/RNN-T & LRS3,VC2,AVS & 
\hline
\end{tabular}
\label{table:hybrid}
\vspace{-0.5cm}
\end{table}

\subsection{Unimodal Inference}

We study the effect of using InterCTC~\cite{lee2021intermediate} and modality dropout~\cite{shi2022learning} during training on unimodal inference results. Table~\ref{table:mod_drop} shows the LRS3 test WER when masking one of the two modalities in the fusion module at test time. We find that training without InterCTC nor modality dropout leads the model to ignore the visual modality. ASR being a substantially easier task than lip reading, the model does not need to learn visual representations to achieve good performance. Forcing the visual encoder to learn hidden representations using one of these two methods solves the problem. We find that solely using modality dropout during training is sufficient to avoid this issue and allows the model to be used in audio-only mode at test time with a minimum loss of performance.

\begin{table}[ht]
\vspace{-0.4cm}
\centering
\scriptsize
\caption{Effect of using InterCTC and/or modality dropout during training on unimodal inference performance.}
\hfill \break
\begin{tabular}{l|lccc}
\hline
\multirow{2}{*}{\textbf{Method}} & \multicolumn{1}{c}{\multirow{2}{*}{\begin{tabular}[c]{@{}c@{}}\textbf{Training} \\ \textbf{Datasets}\end{tabular}}} & \multicolumn{3}{c}{\textbf{AV model LRS3 test WER (\%)}}\\ 
& & \textbf{no mask} & \textbf{mask video} & \textbf{mask audio} \\
\hline\hline
Fast Conformer & LRS3 & 1.9 & 2.0 & 100.0 \\
~~~+ InterCTC & LRS3 & 1.6 & 10.2 & 48.7 \\
~~~+ mod drop & LRS3 & 1.7 & 1.8 & 44.0 \\
~~~~~~+ InterCTC & LRS3 & 1.7 & 1.9 & 44.0 \\
~~~+ mod drop & LRS3,VC2,AVS & 0.8 & 1.0 & 32.2 \\
\hline
\end{tabular}
\label{table:mod_drop}
\vspace{-0.5cm}
\end{table}

\section{Conclusion}

In this work, we presented a multilingual AVSR model using several enhancements to improve performance and audio noise robustness. We adapted the recently proposed Fast Conformer model to process both audio and visual modalities using a novel hybrid CTC/RNN-T architecture. We increased the amount of audio-visual training data for six distinct languages by generating automatic transcriptions of large scale multilingual datasets. Our hybrid audio-visual Fast Conformer model achieved new state-of-the-art results on the LRS3 dataset, reaching WER of 0.8\%. On the recently released MuAViC benchmark, our model yields an absolute average-WER reduction of 11.9\% compared to the original baseline. Models and training recipes will be open-sourced through the NVIDIA NeMo toolkit~\cite{kuchaiev2019nemo}.

\vfill\pagebreak

\label{sec:refs}

{\small
\bibliographystyle{IEEEbib}
\bibliography{strings,refs}
}

\end{document}